\documentclass[12pt,a4]{article}
\usepackage{latexsym,epsfig,amssymb,euscript,amsmath,verbatim,cite, subfigure}
\usepackage{float}
\usepackage{verbatim} 
\newcommand{\be}{\begin{equation}}
\newcommand{\ee}{\end{equation}}
\newcommand{\bea}{\begin{eqnarray}}
\newcommand{\eea}{\end{eqnarray}}

\numberwithin{equation}{section}
\usepackage{bm}
\usepackage{hyperref}

\usepackage[usenames,dvipsnames]{color}

\begin{document}
\pagestyle{empty}
\vspace{1.8cm}

\begin{center}
{\LARGE{\bf{Magneto-transport from momentum dissipating holography}}}

\vspace{1cm}

{\large{Andrea Amoretti$^{a,c,}$\footnote{\tt andrea.amoretti@ge.infn.it },
Daniele Musso$^{b,}$\footnote{\tt dmusso@ictp.it} 
\\[1cm]}}

{\small{
{}$^a$  Dipartimento di Fisica, Universit\`a di Genova,\\
via Dodecaneso 33, I-16146, Genova, Italy\\and\\I.N.F.N. - Sezione di Genova\\
{}$^b$ Abdus Salam International Centre for Theoretical Physics (ICTP)\\
Strada Costiera 11, I 34151 Trieste, Italy\\
{}$^c$  Lorentz Institute for Theoretical Physics, Niels Bohrweg 2, Leiden NL-2333 CA, The Netherlands
}}
\vspace{1cm}

{\bf Abstract}

We obtain explicit expressions for the thermoelectric transport coefficients of a strongly
coupled, planar medium in the presence of an orthogonal magnetic field and momentum-dissipating processes.
The computations are performed within the gauge/gravity framework 
where the momentum dissipation mechanism is introduced by including a mass term for the bulk graviton.
Relying on the structure of the computed transport coefficients and promoting the parameters 
to become dynamical functions, we propose a holography inspired phenomenology
open to a direct comparison with experimental data from the cuprates.

\end{center}

\newpage

\setcounter{page}{1} \pagestyle{plain} \renewcommand{\thefootnote}{\arabic{footnote}} \setcounter{footnote}{0}

\tableofcontents

\section{Introduction and main results}
\label{intro}

Holographic models featuring massive bulk gravitons constitute a promising
framework to obtain relevant information on strongly coupled thermodynamical 
and transport properties of systems which do not conserve momentum. This has a direct impact 
on condensed matter studies where disorder, lattices or heavy degrees of freedom exchange 
momentum with the light degrees of freedom responsible for transport.

We follow the gauge/gravity approach to unveil the structure of thermoelectric transport
in planar, strongly coupled systems in the presence of an orthogonal magnetic field and 
momentum dissipating devices. This structure could be valid also beyond the holographic framework
and therefore furnish general insight or even precise experimental hints.

It is well known on general grounds that holographic models at 
sufficiently high temperature admit a hydrodynamical description \cite{Policastro:2002se}%
\footnote{The possibility of a hydrodynamic analysis relies on a hierarchy of scales such that 
the typical frequency and wave-vector of the studied physics are well below some ultraviolet scale $\Lambda_{\text{UV}}$.
In the present context, the high-temperature regime (although not necessary) provides circumstances where the hydrodynamic assumptions are met.}. 
Therefore, in order to analyze the structure of the transport coefficients we are interested in, 
we can follow the inspiring hint coming from the hydrodynamic analysis performed in \cite{Hartnoll:2007ih}.
There it was shown that the thermoelectric transport properties of a hydrodynamic system 
at non-zero charge density, in the presence of some generic mechanism of momentum dissipation and 
immersed in an external magnetic field are completely determined in terms of the 
thermodynamical properties, the momentum dissipation rate $\tau^{-1}$ and the conductivity at zero net heat current $\tilde\sigma$.
We show explicitly that for the  
holographic models that we study this fact remains true also outside the regime of validity of hydrodynamics
once additional subtleties concerning the definition of a characteristic time for momentum dissipation are taken into account.
In this sense, our results extend the previous hydrodynamical analyses.

To convey the idea of the general aim at stake, it is instructive to think 
of the universal shear viscosity over entropy density ratio $\eta/{\cal S}$.
This is surely one of the most 
important holographic results in momentum conserving models \cite{Kovtun:2004de,Policastro:2001yc}. 
Its universality motivates us to pursue the quests for its counterparts within a more articulate framework where specific momentum dissipating 
phenomena tend to obscure or even impede universal claims. To be more specific, in the absence of momentum 
conservation, one cannot think in terms of the viscosity \cite{Read}, however diffusion bounds for conserved quantities
(e.g. charge and heat) analogous to that concerning $\eta/{\cal S}$ can be possibly formulated \cite{Hartnoll:2014lpa}
(see also \cite{Amoretti:2014ola} for an investigation on the existence of these bounds in holographic models).
The general idea of attempting to tackle the strange metal puzzles relying 
on universal results constitute our main background motivation.

The first main result of the paper consists in obtaining explicit expressions for all the thermoelectric 
transport coefficients in the model-independent language already advocated, namely in terms of 
thermodynamical quantities, a characteristic conductivity $\tilde\sigma$ and a quantity $h$ (to be defined later) 
which depends on the thermodynamic variables ${\cal E}$ and $P$ (i.e. energy density and pressure)
and, in specific regimes, it can be connected to the momentum dissipation rate $\tau^{-1}$. Since this last point is both 
important and delicate, we postpone its detailed treatment to later sections; for the moment being, it is possible to regard $h$ 
as a convenient bookkeeping device to express compactly the transport coefficients in view of studying their scaling properties.

The explicit expressions that we obtain for the transport coefficients are
\begin{align}\label{aspuni}
&\sigma_{xx}= h\ \frac{  \rho^2 + \tilde\sigma \left(B^2 \tilde\sigma + h\right) }
{B^2 \rho^2 + \left(B^2 \tilde\sigma + h\right)^2 } \ , 
&\sigma_{xy}= \rho B\ \frac{\rho^2 + \tilde\sigma \left(B^2 \tilde\sigma + 2\, h\right)}
{B^2 \rho ^2+\left(B^2 \tilde\sigma + h \right)^2} \ , \\\label{aspuni2}
&\alpha_{xx}= \frac{\rho\,  \mathcal{S}\ h}{B^2 \rho ^2+\left(B^2 \tilde\sigma+ h \right)^2} \ , 
&\alpha_{xy}= {\cal S} B \ \frac{\rho^2 + \tilde\sigma \left(B^2 \tilde\sigma + h\right)}
{B^2\rho ^2+\left(B^2 \tilde\sigma+ h \right)^2} \ , \\\label{aspuni3}
&\bar{\kappa}_{xx}=\frac{ \mathcal{S}^2 T \left( B^2 \tilde\sigma + h \right)}
{ B^2 \rho ^2+\left(B^2   \tilde\sigma + h \right)^2 } \ , 
&\bar{\kappa}_{xy}=\frac{B \rho\,  \mathcal{S}^2 T}
 { B^2 \rho^2+\left(B^2 \tilde\sigma + h \right)^2 } \ .
\end{align}
The complete derivation of these formul\ae~and further comments are given in later sections.
Here we just observe that they are suitable to be compared to and/or integrated with phenomenological data.
On this observation we build our second main result. 
 Namely we investigate the possibility of considering the transport coefficients 
\eqref{aspuni}, \eqref{aspuni2}, \eqref{aspuni3} independently of their specific holographic origin;
more precisely, we allow for phenomenological properties (e.g. temperature scalings) to be
implemented in the various quantities appearing in the transport coefficients.
This possibility clearly departs from the holographic computation and promotes the structure of the 
transport coefficients as a new starting point of a phenomenological analysis;
it is essential to stress that the extra freedom we allow ourselves is justified by the
direct exposure to experimental data as we will illustrate carefully.
A completely analogous approach (which here is extended to the whole set of thermo-electric
transport coefficients) inspired the analysis of \cite{Blake:2014yla}.

In addition to the comparison with experiments,
there remains the interesting open problem whether the phenomenology we describe could be precisely embedded in enlarged holographic models;
technically this concerns the possibility of keeping under control all the necessary complications of promoting the model parameters to become,
in some sense to be specified, dynamical functions.
Without pretending to reach this goal now, we can however start to check if this extra freedom we allow ourselves 
(even before motivating it on conclusive theoretical grounds) could nevertheless permit to encompass the experimental data with few assumptions.

The paper is structured as follows. In Section \ref{model} we first introduce the holographic model proposed
in \cite{Vegh:2013sk} and describe the dyonic black brane solutions needed to account for a boundary 
theory immersed in a magnetic field. Sections \ref{thermodynamics} and \ref{tran} constitute the computational core 
of the paper and they are dedicated to the analysis of thermodynamical and transport properties respectively.
The reader interested in the physical outcome (in spite of the specific computational details) could jump 
directly to Section \ref{uni} where we set the arguments for the possible applicability of the formul\ae~for the transport coefficients 
beyond the particular setup where they were obtained. Specifically, we compare and extend the hydrodynamical 
analysis performed in \cite{Hartnoll:2007ih} with careful attention paid to the electromagnetic duality 
of the bulk model and its effects to the boundary theory. In Section \ref{pheno} we make contact with the phenomenology;
namely we supplement the holographic model with phenomenological information and compare the resulting  
set of scalings for the transport coefficients with the experimental literature. Eventually in Section \ref{discu}
we give our concluding remarks suggesting many interesting directions of further investigation.
Some computational details are given in the Appendix.

\section{The holographic model}
\label{model}

The model we consider features massive gravitons in the bulk. Its first consideration
within the holographic context is due to \cite{Vegh:2013sk} and then it has been further analyzed in 
\cite{Davison:2013jba,Blake:2013bqa,Blake:2014yla,Amoretti:2014zha,Amoretti:2014mma,Amoretti:2014ola}.
Since we are interested in describing a magnetic field, we generalize the dyonic black hole holographic 
models \cite{Hartnoll:2007ai,Hartnoll:2007ip} to include massive gravity.

The gravitational action is
\begin{equation}
\label{massivelag}
\begin{split}
S = \int d^4x\ & \sqrt{-g} \left\{ \frac{1}{2 \kappa_4^2}\left[R+\frac{6}{L^2}+\beta \left([\mathcal{K}]^2
-[\mathcal{K}^2]\right) \right]-\frac{1}{4 q^2}F_{\mu \nu}F^{\mu \nu} \right\}\\
&+\frac{1}{2 \kappa_4^2} \int_{z=z_{\text{UV}}} d^3 x\ \sqrt{-g_b} \  2 K  \ ,
\end{split}
\end{equation}
where $\beta$ is a parameter having the dimension of a mass squared, $L$ is the AdS$_4$ radius, 
$\kappa_4$ is the gravitational constant in $3+1$ spacetime dimensions and $F_{\mu \nu} \equiv \partial_{[\mu} A_{\nu]}$ 
is the field strength of the gauge field $A_{\mu}$. We also supplemented the bulk action 
with the usual Gibbons-Hawking counter-term, expressed in terms of the metric $(g_b)_{\mu \nu}$ induced on the radial shell corresponding to the UV cut-off
 and the trace $K$ of the extrinsic curvature $K_{\mu \nu}$ of the same manifold%
\footnote{The UV cut-off $z_{\text{UV}}$ is eventually sent to zero in the final step of the holographic renormalization procedure
which we describe later.}. The Gibbons-Hawking counter-term is necessary to have a well defined bulk variational problem.

The mass term for the gravitons is proportional to $\beta$ and it is 
defined in terms of the trace (indicated with small square brackets in \eqref{massivelag}) 
of the two matrices $\mathcal{K}^{\mu}_{\;\nu}$ and $\left(\mathcal{K}^2\right)^{\mu}_{\;\nu}$ defined as follows
\begin{equation}
\begin{split}
&(\mathcal{K}^2)^\mu_{\ \nu}\equiv g^{\mu \rho}f_{\rho \nu} \ , \qquad \mathcal{K}^\mu_{\ \nu} \equiv \left( \sqrt{ \mathcal{K}^2} \right)^{\; \mu}_{\; \; \;\nu}\ , 
\end{split}
\end{equation}
where $f_{\mu \nu}$ is a fixed, non-dynamical metric. This fiducial metric $f_{\mu \nu}$ controls the way in which the 
graviton mass potential breaks the diffeomorphism invariance.
Since we want to discuss bulk solutions dual to isotropic and homogeneous configurations of 
the boundary field theory where momentum is non-conserved, we need to consider the fiducial metric 
\begin{equation}
f_{\mu \nu}=\text{diag}\{0,0,1,1\} \ ,
\end{equation}
as described in \cite{Vegh:2013sk}.
Indeed, the non-trivial diagonal entries correspond to the boundary spatial directions. 
The trivial ones instead correspond to the radial and time directions along which we want to preserve 
bulk diffeomorphism invariance. The former being justified by the adoption of the standard holographic 
approach itself (where the holographic coordinate parametrizes the boundary theory renormalization flow),
the latter being associated to energy conservation and, technically, to the construction of thermal
solutions and thermal fluctuations in terms of the metric fields (see \cite{Hartnoll:2009sz,Herzog:2009xv,Musso:2014efa}).
We do not discuss here the stability of the massive gravity model and questions related to ghost modes 
for which we refer to \cite{Vegh:2013sk,Alberte:2014bua} and references therein.
As a general comment, we underline the bottom-up nature of the holographic model at hand 
and the associated analysis; said otherwise, the present study belongs to the (today already wide and lively) investigation line
adopting massive gravity holography in analogy to its massless counterparts even though interesting and foundational questions such as a
conclusive definition of the holographic dictionary or a UV embedding of massive gravity remain open.

As studied in \cite{Blake:2013owa}, there is an important connection between massive gravity 
and setups featuring spatially modulated sources; such a connection can be made quantitative comparing the structure of the bulk equations of motion
arising in massive gravity to that arising in a model featuring a (weak) modulated source for a neutral scalar (see \cite{Blake:2013owa} for details). 
Already before a conclusive derivation of massive gravity from microscopic models breaking translation invariance,
the analysis of \cite{Blake:2013owa} hints to massive gravity as a valuable effective and holographic approach to account for an explicit breaking of momentum conservation
suitable to describe a non-dynamical lattice or the presence of quench disorder (see \cite{Davison:2013txa}) and heavy impurities%
\footnote{Along similar lines, it is relevant to mention that the analyses involving explicit spatially dependent 
sources allow also for a direct implementation of disorder in holography \cite{Arean:2013mta}.}.

We want to discuss the effects due to the presence of an external magnetic field $B$ 
orthogonal to the plane $xy$. In particular its consequences on the thermoelectric transport coefficients
in the holographic system \eqref{massivelag} at non-zero chemical potential $\mu$.
To include the constant magnetic field $B$ we adopt the following ansatz for the background metric $g_{\mu \nu}$ and the gauge field $A_{\mu}$
\begin{equation}\label{bbm}
\begin{split}
&ds^2=\frac{L^2}{z^2} \left[-f(z) dt^2 + dx^2 + dy^2 + \frac{1}{f(z)} dz^2\right]\ ,\\
& A=\phi(z)\, dt + B\, x\, dy\ .
\end{split}
\end{equation}
Substituting this ansatz within the equations of motion derived from \eqref{massivelag}, we obtain the following black brane solution
\begin{equation}
\label{solublack}
\begin{split}
&\phi(z)= \mu - q^2 \rho z = \mu \left(1-\frac{z}{z_h} \right) \ , \qquad \rho \equiv \frac{\mu}{q^2 z_h} \ , \\
&f(z)=1-\frac{z^3}{z_h^3}
      +\beta \left(z^2-\frac{z^3}{z_h}\right)
   -\frac{z^3 }{z_h} \left(1-\frac{z}{z_h}\right) \frac{\gamma ^2  \left(B^2 z_h^2+\mu ^2\right)}{2 L^2}\ ,
\end{split}
\end{equation}
where we have denoted with $z_h$ the horizon location defined by the vanishing of the 
emblackening factor, namely $f(z_h)=0$.
The definition of $\rho$ is actually substantiated by the explicit analysis of the 
thermodynamics that we perform in Section \ref{thermodynamics}. For the sake of practical 
convenience, we introduced $\gamma \equiv \kappa_4 / q$.

\section{Thermodynamics}
\label{thermodynamics}

As discussed in the previous section, the black brane solution \eqref{solublack} corresponds to a 
planar dyonic black hole having both electric and magnetic charges. From the boundary theory standpoint, $B$ represents 
a magnetic field perpendicular to the spatial manifold $xy$ which enters the boundary thermodynamical quantities;
as usual in gauge/gravity, these are derived from the bulk on-shell action as we now show in detail.

The temperature $T$ and the entropy density $\mathcal{S}$ are the easiest thermodynamical quantities to compute
since they are determined from the horizon data, namely
\begin{equation}
\label{tempentr}
T = -\frac{f'(z_h)}{4 \pi}=-\frac{\kappa_4 ^2 z_h^2 \left(B^2 z_h^2+\mu ^2\right)-2 L^2 q^2 \left(\beta  z_h^2+3\right)}{8 \pi  L^2 q^2 z_h} \ ,
\ \ \ \ \ \
\mathcal{S} = \frac{2 \pi L^2}{\kappa_4^2 z_h^2} \ .
\end{equation}

In order to compute the energy density $\mathcal{E}$, the pressure $P$, the charge density $\rho$ and the magnetization $M$,
we need to evaluate explicitly the Landau potential $\Omega$ which, according to the holographic dictionary,
is identified with the on-shell bulk action. Not surprisingly, the bulk action \eqref{massivelag} when naively evaluated on the solution 
\eqref{solublack} is divergent and therefore needs to be renormalized. The standard renormalization process 
consists in regularizing the action by means of a UV cut-off $z_{\text{UV}}$ and supplementing it with appropriate counter-terms.
These could necessarily be written in terms of boundary fields but (proceeding as in \cite{Blake:2013bqa}),
once evaluated on the solution \eqref{solublack}, can be explicitly expressed as follows
\begin{equation}
S_{\text{ct}}=\frac{1}{2 \kappa_4^2}\int_{z=z_{\text{UV}}} \sqrt{-g_b}\left(\frac{4}{L}+ \frac{2}{L}\beta z_{\text{UV}}^2 \right) \ ,
\end{equation}
where $g_b$ is the metric induced on the $z=z_{\text{UV}}$ shell.
With this counter-term in place, the Landau potential $\Omega$ assumes the following form
\begin{equation}
\Omega \equiv \lim_{z_{\text{UV}} \rightarrow 0} \left( S+S_{\text{ct}} \right)_{\text{on-shell}}=V \left(\frac{3 B^2 z_h}{4 q^2}-\frac{L^2}{2 \kappa ^2
   z_h^3}+\frac{\beta  L^2}{2 \kappa ^2 z_h}-\frac{\mu ^2}{4 q^2 z_h}\right) \ .
\end{equation}
We have denoted with $V$ the boundary spatial volume.

Once the Landau potential is known, the other thermodynamical quantities follow easily
by means of standard thermodynamical relations. We explicitly obtain%
\footnote{In order to compute the thermodynamical derivatives with respect to $T$, $\mu$ and $B$ 
one must recall that $z_h$ is an implicit function of these quantities as given in \eqref{tempentr}.}
\begin{equation}
P=-\frac{\Omega}{V}=-\frac{3 B^2 z_h}{4 q^2}+\frac{L^2}{2 \kappa_4 ^2 z_h^3}-\frac{\beta  L^2}{2 \kappa_4 ^2 z_h}+\frac{\mu ^2}{4 q^2
   z_h} \ ,
\end{equation}
\begin{equation}
\mathcal{E}=-P+\mathcal{S}T+\mu \rho=\frac{B^2 z_h}{2 q^2}+\frac{L^2}{\kappa_4 ^2 z_h^3}+\frac{\beta  L^2}{\kappa_4 ^2 z_h}+\frac{\mu ^2}{2 q^2 z_h}\ ,
\end{equation}
\begin{equation}
 \rho = \frac{\partial \mathcal{E}}{\partial \mu} =\frac{\mu}{q^2 z_h} \ , \qquad  M = - \frac{\partial \mathcal{E}}{\partial B} = -\frac{B z_h}{q^2} \ .
\end{equation}

For later purposes it is useful to define one additional and less common
thermodynamical quantity, namely the magnetization energy $M^E$ defined as 
\begin{equation}
M^E \equiv- \frac{\delta \Omega}{\delta F_{xy}^E} \ ,
\end{equation}
where $F_{xy}^E=\partial_x \delta g_{ty}-\partial_y \delta g_{tx}$ and $\delta g_{ta}$ represents a variation of the metric sourcing $T^{ta}$. 
Following \cite{Hartnoll:2007ih}, an operative method to evaluate the magnetization energy consists in computing the on-shell action on the following solution
\begin{equation}\label{trick}
\begin{split}
&A_t=\phi(z) \ , \qquad A_y=Bx-\left(\phi(z)-\mu\right)B^Ex \ , \\
ds^2=\frac{L^2}{z^2} &\left[-f(z) \left(dt-B^E x dy \right)^2 + dx^2 + dy^2 + \frac{1}{f(z)} dz^2\right]\ \ ,
\end{split}
\end{equation}
where $\phi(z)$ and $f(z)$ are the same as in \eqref{solublack};
the magnetization energy $M^E$ is then obtained by differentiating the on-shell action with respect to $B^E$ and finally setting $B^E$ to zero
as one can understand comparing de definition of $F_{xy}^E$ and the explicit form for the metric in \eqref{trick}.
This computation proceeds exactly in the same way as in \cite{Hartnoll:2007ih} and the mass term for the graviton does not affect the final result. 
Eventually we obtain
\begin{equation}
\label{ME}
M^E=\frac{\mu M}{2} \ .
\end{equation}

\section{Transport coefficients}
\label{tran}

 We compute analytically the whole set of thermoelectric DC transport coefficients for the boundary
theory corresponding to the bulk model \eqref{massivelag}. To this aim, we employ the method
first illustrated in \cite{Donos:2014cya} and subsequently applied in \cite{Amoretti:2014mma,Blake:2014yla,Donos:2014yya,Cheng:2014tya,Donos:2014gya}.
Such an approach enforces and extends the so-called ``membrane paradigm'' \cite{Iqbal:2008by} to momentum-dissipating systems
and relies on the analysis of quantities which do not evolve along the holographic direction from the IR to the UV.
 
\subsection{Electric conductivity}
\label{elcon}

We consider linearized fluctuations around the bulk background solution \eqref{solublack}.
Following \cite{Donos:2014cya}, to the purpose of computing the linear response to a ``pure'' electric field
(i.e. in the absence of a thermal gradient), one considers the following ansatz for the fluctuating fields
\begin{eqnarray}\label{ansaflu}
 a_i(t,z) &=& - E_i\, t + \tilde{a}_i(z)\ ,\\
 h_{ti}(t,z) &=& \tilde{h}_{ti}(z)\ ,\\
 h_{zi}(t,z) &=& \tilde{h}_{zi}(z)\ ,
\end{eqnarray}
where $i = x,y$; we henceforth adopt small Latin letters to refer to spatial boundary indices%
\footnote{Note that the magnetic field mixes the $x$ and $y$ fluctuation sectors
and therefore all the components along these directions in \eqref{ansaflu} must be switched on.}.
The vector $E_i$ introduced in the ansatz corresponds to an external electric field perturbing 
the system.

The quantity
\begin{equation}\label{curre}
 \bar{J}^\mu = - \frac{\sqrt{-g}}{q^2}\, F^{z\mu}\ ,
\end{equation}
(where $\mu = t,x,y$ is a boundary spacetime index) is conserved along the holographic direction
as a direct consequence of the Maxwell equation for the fluctuations. Indeed,
recalling the ansatz \eqref{ansaflu}, we obtain
\begin{equation}\label{consa}
 \sqrt{-g}\,  \nabla_M F^{MN} = 0\ \ \ \ \longrightarrow\ \ \ \ \partial_z (\sqrt{-g} F^{zi}) = 0\ .
\end{equation}
The capital indices refer to the bulk spacetime and the arrow 
means that we consider just the spatial components.
The quantities  $\bar{J}^i$ are radially conserved and explicitly given by
\begin{equation}\label{barJ}
 \bar{J}_i = -\frac{f(z)}{q^2}\, \tilde{a}_i'(z)
 -\frac{B z^2 f(z)}{L^2 q^2}\, \epsilon_{ij}\, \tilde{h}_{zj}(z)
 +\frac{z^2 \mu}{L^2 q^2 z_h}\, \tilde{h}_{ti}(z)\ ,
\end{equation}
where $\epsilon_{ij} = -\epsilon_{ji} = 1$. To obtain \eqref{barJ} we
have again referred to the ansatz \eqref{ansaflu} and considered just up to the linear order in the 
fluctuating fields. We remind the reader that the boundary indices are raised 
and lowered with the flat boundary Minkowski metric.

To study the near-horizon behavior of the fluctuating fields
and demand regular infrared behavior, it is convenient to adopt the Eddington-Finkelstein coordinates, namely
\begin{equation}
 v = t - \frac{1}{4 \pi T}\, \log \left[\frac{z_h-z}{L}\right] \ ,
\end{equation}
leaving all the other bulk coordinates untouched. Skipping the details (which can be found in the analogous 
computation described in \cite{Amoretti:2014mma}), the infrared regularity requirement amounts to having 
the following asymptotic behaviors

\begin{equation}\label{hti_hor}
 \tilde{h}_{ti}(z) = f'(z_h)\, \tilde{h}_{zi}(z_h) (z_h-z) + \mathcal{O}(z_h-z)^2 \ ,
\end{equation}
and
\begin{equation}\label{a_hor}
 \tilde{a}_i(z) =  \frac{E_i}{4 \pi T}\, \log \left[\frac{z_h-z}{L}\right] + \mathcal{O}(z_h-z)\ .
\end{equation}
To avoid clutter, we relegated the explicit expressions of the equations of motion 
for the linearized fluctuations in Appendix \ref{El_app}. It is however important to recall that 
$\tilde{h}_{zi}(z)$ is governed by an algebraic equation and therefore expressible in terms of the 
other fluctuating fields (hence it does not demand further IR requirements%
\footnote{The presence of an algebraic equation for a fluctuation indicates a residual
gauge invariance of the fluctuation system; one can indeed solve the algebraic equation and from 
then on ``forget'' about the solved fluctuation. For related comments see \cite{Donos:2014gya,Kim:2015sma}.}). An explicit
infrared asymptotic analysis returns
\begin{equation}\label{hti_val}
 \tilde{h}_{ti}(z_h) = - \frac{L^2 \gamma^2 \left[B \epsilon_{ij} E_j z_h \left(\gamma^2 B^2 z_h^2 + \gamma^2 \mu^2 -L^2 \beta \right) + E_i L^2 \beta  \mu\right]}
 {z_h \left(L^2 \beta -B^2 z_h^2 \gamma ^2\right)^2+B^2 z_h^3 \gamma ^4 \mu ^2}\ ,
\end{equation}
which we report explicitly for the sake of completeness and to underline that its $B\rightarrow 0$ limit is consistent
with previous results obtained directly at $B=0$ in \cite{Amoretti:2014mma}. 

It is essential to observe that, turning the attention to the 
near-boundary asymptotics, one actually identifies $\bar{J}(z=0)$ with the electric current $J_i$
of the boundary theory,
\begin{equation}
 \bar{J}_i(z=0)=J_i \ .
\end{equation}
Moreover, being $\bar{J}$ radially conserved, it can be evaluated in the IR and then
expressed exclusively in terms of the near-horizon asymptotic data, namely $\bar{J}(z=0)=\bar{J}(z=z_h)$.

The electric conductivity matrix is
\begin{equation}
 J_i = \sigma_{ij} E_j\ .
\end{equation}
Hence its entries are directly read from the explicit expression of the electric current;
this yields
\begin{equation}
\sigma_{xx}=\frac{\beta  L^2 \left[\beta  L^2 q^2-\kappa_4 ^2 \left(B^2 z_h^2+\mu ^2\right)\right]}{-2 \beta  B^2 \kappa_4 ^2 L^2 q^2
   z_h^2+B^2 \kappa_4 ^4 z_h^2 \left(B^2 z_h^2+\mu ^2\right)+\beta ^2 L^4 q^4}\ ,
\end{equation}
and
\begin{equation}
\sigma_{xy}=\frac{B \mu  z_h \left[\kappa_4 ^4 \left(B^2 z_h^2+\mu ^2\right)-2 \beta  \kappa_4 ^2 L^2 q^2\right]}{q^2 \left[-2 \beta  B^2
   \kappa_4 ^2 L^2 q^2 z_h^2+B^2 \kappa_4 ^4 z_h^2 \left(B^2 z_h^2+\mu ^2\right)+\beta ^2 L^4 q^4\right]}\ .
\end{equation}
We have $\sigma_{xx} = \sigma_{yy}$ and $\sigma_{xy} = -\sigma_{yx}$.

\subsection{Thermoelectric response}
\label{resp}

We want now to compute the thermoelectric conductivities $\alpha_{xx}$ and $\alpha_{xy}$. 
Considering the system at non-zero electric field and zero thermal gradient, 
these two conductivities are defined by the following relation
\begin{equation}
\label{defthermel}
Q_i=\alpha_{ij}\, T\,  E^j \ ,
\end{equation}
where $Q^i$ is the heat current in the $i$-direction, which can be related to the boundary 
stress-energy tensor $T^{\mu \nu}$ and to the electric current $J^{\mu}$ by the identity $Q^i=T^{ti}-\mu J^{i}$.

In order to apply the same procedure used for the electric conductivity
in the previous section, we need to define in the gravitational system \eqref{massivelag} 
a quantity $\bar{Q}_i$ which is radially conserved and that, choosing the appropriate UV 
boundary conditions for the functions appearing in the ansatz (see \cite{Donos:2014cya,Amoretti:2014mma}), 
can be identified at the boundary with the heat current $Q_i$. 
As illustrated in \cite{Donos:2014cya}, the quantity which does the game in the absence of an external magnetic field is:
\begin{equation}
\label{qbar1}
\bar{Q}^i_1=\frac{\sqrt{-g}}{\kappa_4^2}\, \nabla^z k^i - \phi(z) \, \bar{J}^i \ ,
\end{equation}
where $k=\partial_t$. The proof that this quantity is radially conserved relies on the fact that $k$ is a 
Killing vector for the gravitational action \eqref{massivelag} (see \cite{Donos:2014cya} for more details). 

When one considers circumstances with a non-zero magnetic field $B$, the quantity \eqref{qbar1} is no longer radially conserved. 
In fact, differentiating $\bar{Q}^i$ with respect to $z$ and evaluating the result on the equations of motion (see Appendix \ref{El_app}) we obtain
\begin{equation}
\partial_z \bar{Q}^i_1=\epsilon^{ij}E_j\frac{B }{q^2} \ .
\end{equation}
We can therefore define the following radially conserved quantity,
\begin{equation}
\label{qbar}
 \bar{Q}^i=\frac{\sqrt{-g}}{\kappa_4^2}\, \nabla^z k^i - \phi(z) \, \bar{J}^i-\epsilon^{ij}E_j\frac{B }{q^2}z \ .
\end{equation}
Evaluating the expression \eqref{qbar} on the ansatz \eqref{ansaflu} at the linear order in the fluctuations we obtain
\begin{equation}\label{Qi}
\begin{split}
\bar{Q}_i = \frac{f(z) \phi (z)}{q^2}\, \tilde{a}_i'(z) &
-\frac{B z}{q^2}\, \epsilon_{ij} E_j 
+\frac{B z^2 f(z) \phi (z)}{L^2 q^2} \, \epsilon_{ij} \tilde{h}_{zj}(z) \\
&+\left(-\frac{f'(z)}{2 \kappa_4 ^2}+\frac{f(z)}{z \kappa_4 ^2}-\frac{z^2 \mu\,  \phi (z)}{L^2 q^2 z_h}\right) \tilde{h}_{ti}(z)
+\frac{f(z)}{2\kappa_4 ^2}\, \tilde{h}_{ti}'(z)\ .
\end{split}
\end{equation}
The proof that $\bar Q$ corresponds to the heat current at the boundary is straightforward. 
In fact, the third term in \eqref{qbar} vanishes at $z=0$, the second term is equal%
\footnote{We remind the reader that the quantity $\bar{J}^i$ is radially conserved and coincides with the boundary 
electric current, as illustrated in the previous section.} to $-\mu J^i$, 
and the first term coincides at the boundary with the $ti$ component of the holographic stress-energy tensor, 
\begin{equation}\label{Tti}
 T^{ti} = \frac{L^5}{\kappa_4^2 z^5} \left(-K^{ti} + K g^{ti}_b + \frac{2}{L}g^{ti}_b\right)
        =\frac{\tilde{h}_{ti}'(z)}{2 \kappa_4 ^2 \sqrt{f(z)}}-\frac{\tilde{h}_{ti}(z)}{z \kappa_4^2 \sqrt{f(z)}}+\frac{2\, \tilde{h}_{ti}(z)}{z \kappa_4 ^2 f(z)} \ .
\end{equation}

Exploiting its radial conservation, we can compute $\bar{Q}^i$ at $z=z_h$
and express the heat current only in terms of horizon data. Finally, using the definition of the thermoelectric conductivities \eqref{defthermel}, we obtain:
\begin{equation}
\alpha_{xx}=-\frac{2 \pi  \beta  \mu  L^4 q^2}{-2 \beta  B^2 \kappa ^2 L^2 q^2 z_h^3+B^2 \kappa ^4 z_h^3 \left(B^2 z_h^2+\mu
   ^2\right)+\beta ^2 L^4 q^4 z_h} \ ,
\end{equation}
and
\begin{multline}
\alpha_{xy}=
\frac{2 \pi  B L^2 \left(\kappa_4 ^2 \left(B^2 z_h^2+\mu ^2\right)-\beta  L^2 q^2\right)}{-2 \beta  B^2 \kappa_4 ^2 L^2 q^2
   z_h^2+B^2 \kappa_4 ^4 z_h^2 \left(B^2 z_h^2+\mu ^2\right)+\beta ^2 L^4 q^4}\\ +\frac{8 B \pi  L^2 q^2 z_h^2}{\kappa_4 ^2q^2 z_h^2 \left(B^2 z_h^2+\mu ^2\right)-2 L^2 q^4 \left(\beta  z_h^2+3\right)} \ .
\end{multline}
However, as illustrated in \cite{Hartnoll:2007ih,cooper}, in order to properly define the thermoelectric response
in the presence of a magnetic field, 
one has to subtract to the heat current the contribution due to the magnetization current%
\footnote{We note that the radially conserved quantity \eqref{qbar1} is defined up to an additive constant. 
Therefore it would be possible to add to $\bar{Q}_1^i$ the constant $\epsilon_{ij}E_j \frac{B z_h}{q^2}$
and the radially conserved quantity defined in such a way would coincide with the magnetization-subtracted heat current.
This is in line with the analysis of \cite{Blake:2015ina}.}.
This implies that the off-diagonal components of the thermoelectric conductivity have to be defined as
\begin{equation}
\alpha_{xy}^{\text{sub}}=\alpha_{xy}+\frac{M}{T} \ ,
\end{equation}
and, recalling the explicit expression of the temperature $T$ and of the magnetization $M$ derived in Section \ref{thermodynamics}, we obtain
\begin{equation}
\alpha_{xy}^{\text{sub}}=\frac{2 \pi  B L^2 \left(\kappa_4 ^2 \left(B^2 z_h^2+\mu ^2\right)-\beta  L^2 q^2\right)}{-2 \beta  B^2 \kappa_4 ^2 L^2 q^2
   z_h^2+B^2 \kappa_4 ^4 z_h^2 \left(B^2 z_h^2+\mu ^2\right)+\beta ^2 L^4 q^4} \ .
\end{equation}
From now on, we will refer to $\alpha_{xy}^{\text{sub}}$ as the off-diagonal component of the thermoelectric conductivity and we will 
indicate it with $\alpha_{xy}$.

\subsection{Thermal conductivity}

The thermal conductivities $\bar{\kappa}_{xx}$ and $\bar{\kappa}_{xy}$ are defined in terms of
the heat current generated by the presence of an external thermal gradient $\nabla_i T$ in the following way
\begin{equation}
\label{defkappa}
Q_i=-\bar{\kappa}_{ij}\nabla_j T
\end{equation}
As illustrated in \cite{Donos:2014cya}, in order to study the holographic model \eqref{massivelag}
in the presence of an external thermal gradient and at zero electric field we have to consider the following ansatz for the fields of the theory
\begin{equation}\label{ansaflu2}
 \begin{split}
  &a_i(t,z)    = s_i\, \phi(z)\, t + \tilde{a}_i(z)\ , \\
  &h_{ti}(t,z) = -s_i \frac{L^2}{z^2} f(z)\, t + \tilde{h}_{ti}(z)\ , \\
  &h_{zi}(t,z) = \tilde{h}_{zi}(z) \ ,
 \end{split}
\end{equation}
where $s_i$ can be proven to be equal to the quantity $-\frac{\nabla_i T}{T}$ in the boundary field theory 
(see \cite{Donos:2014cya}). The linearized equations of motion for this ansatz can be found in Appendix \ref{theapp}.

As in the thermo-electric case, the quantity \eqref{qbar1} evaluated on the ansatz \eqref{ansaflu2} is not radially conserved. 
On the other hand, the quantity (already given in \eqref{qbar})
\begin{equation}
\label{Qterm}
\bar{Q}^i=\frac{\sqrt{-g}}{\kappa_4^2}\, \nabla^z k^i - \phi(z) \, \bar{J}^i+\epsilon^{ij}s_j\frac{B \mu}{q^2}z \ ,
\end{equation}
is radially conserved once evaluated on shell.

The computation of the thermal conductivities is now straightforward.  
As in the previous section, we evaluate the radially conserved quantity \eqref{Qterm} on the 
ansatz \eqref{ansaflu2} at the linear order in the fluctuations. Also in this case, as long 
as we consider the DC response, $\bar{Q}_i$ can be proven to coincide with the heat current 
in the boundary field theory (see Appendix \ref{appthermalT} for further details on this point)%
\footnote{Actually, considering the ansatz \eqref{ansaflu2} there are some additional technical 
difficulties in proving this statement due to the fact that the quantity $\nabla^{z} k^{i}$ 
differs from the holographic stress-energy tensor $T^{ti}$ by terms linear in the time coordinate $t$. 
However, as proven in \cite{Donos:2014cya}, these terms do not contribute to DC transport properties.}. 
Finally, computing the quantity $\bar{Q}_i$ at the horizon $z=z_h$, and considering the definition of the thermal conductivities \eqref{defkappa} we obtain
\begin{equation}
\bar{\kappa}_{xx}=\frac{1}{T}\frac{\bar{Q}_x}{\alpha_x}=-\frac{\pi  L^2 \left(\beta  L^2 q^2-B^2 \kappa ^2 z_h^2\right) 
\left[2 L^2 q^2 \left(\beta  z_h^2+3\right)-\kappa ^2
   z_h^2 \left(B^2 z_h^2+\mu ^2\right)\right]}{2 \kappa ^2 z_h^3 \left[-2 \beta  B^2 \kappa ^2 L^2 q^2 z_h^2+B^2
   \kappa ^4 z_h^2 \left(B^2 z_h^2+\mu ^2\right)+\beta ^2 L^4 q^4\right]}
\end{equation}
Also in this case we need to subtract the contribution due to the magnetization current \cite{Hartnoll:2007ih,cooper}
from the off-diagonal conductivity $\bar{\kappa}_{xy}$, namely
\begin{equation}
\bar{\kappa}_{xy}=\frac{1}{T}\frac{\bar{Q}_x}{\alpha_y}+\frac{2\left(M^E-\mu M\right)}{T}=
-\frac{\pi  B \mu  L^2 \left[\kappa_4 ^2 z_h^2 \left(B^2 z_h^2+\mu ^2\right)-2 L^2 q^2 \left(\beta 
   z_h^2+3\right)\right]}{2 z_h^2 \left[-2 \beta  B^2 \kappa_4 ^2 L^2 q^2 z_h^2+B^2 \kappa_4 ^4 z_h^2 \left(B^2
   z_h^2+\mu ^2\right)+\beta ^2 L^4 q^4\right]} \ ,
\end{equation}
where we have used equation \eqref{ME} for $M^E$.

\section{Structure of the thermoelectric transport coefficients}
\label{uni}

The behaviors of the transport coefficients found in the previous 
section depend on the specific form of the thermodynamical quantities obtained in the massive 
gravity model in Section \ref{thermodynamics}.
Nevertheless, as we will show, these transport coefficients can be cast in a form which may aspire to be universal,
at least within the holographic framework%
\footnote{In this and the following sections the attribute ``universal'' must be intended as restricted to 
the class of spatially homogeneous systems. Similarly, throughout the paper, all the relations and explicit expressions connecting the 
transport properties to the thermodynamics refer to spatially homogeneous systems only. We thank the referee for having 
indicated this point.}. It is tempting, in fact, to argue that the 
formul\ae~obtained through holography with massive gravity could remain valid in a more general framework, 
namely as long as one considers strongly coupled systems with neutral mechanisms of momentum dissipation%
\footnote{This statement might not be valid in cases when the momentum dissipation is obtained by 
adding an additional gauge field as in \cite{Donos:2014oha}.} 
(e.g. impurities).
This statement is partially corroborated by two recent papers appeared just a few days after the first version of this manuscript \cite{Blake:2015ina,Lucas:2015pxa,Kim:2015wba}, 
where formul\ae~for the thermo-electric transport coefficients in agreement with those obtained in this paper are computed 
in holographic Q-lattices \cite{Blake:2015ina} and, independently of holography, by means of the memory matrix approach \cite{Lucas:2015pxa}.
 
In order to proceed, we write the full set of transport coefficients computed in massive gravity in terms of 
the thermodynamical quantities and two transport quantities $\tilde\sigma$ and $h$ (as explicitly reported in \eqref{aspuni}, \eqref{aspuni2}, \eqref{aspuni3}). 
Let us start defining the explicit expressions for
these latter in the model at hand. As far as $\tilde\sigma$ is concerned, it appears manifestly from the expressions 
of the transport coefficients \cite{Blake:2013bqa,Amoretti:2014zha,Amoretti:2014mma} that
it represents the conductivity at zero net heat current, then it takes the usual  form $\tilde\sigma=1/q^2$
and it is directly connected to the parameter $q$ of the bulk model.
Regarding $h$, we define it explicitly as follows
\begin{equation}
\label{para}
h = - \frac{\mathcal{S} \beta}{2 \pi } = 
\frac{4 \beta  L^2 q^2   z_h}{\kappa_4 ^2 z_h^2 \left(B^2 z_h^2-3 \mu ^2\right)-2 L^2 q^2 \left(\beta  z_h^2+3\right)} \ ,
\end{equation}
noticing its relation to the parameter $\beta$. 

In order to gain physical intuition on the nature of $h$, we connect it to the 
characteristic time $\tau$ of momentum dissipation at $B=0$ 
\begin{equation}\label{cu}
\tau|_{B=0}=-\frac{2 \pi \left( \mathcal{E}+P \right)}{\mathcal{S} \beta} = - \frac{2 L^2 q^2 \left(\beta  z_h^2+3\right)}{4 \beta  L^2 q^2
   z_h} \ ,
\end{equation}
as found in \cite{Davison:2013jba} through an analysis at low momentum dissipation; combining \eqref{para} and \eqref{cu}, for $h$ we have
\begin{equation}\label{conne}
 \left.h\right|_{B=0} = \frac{\mathcal{E}+P}{\tau}\ .
\end{equation}
We also stress that, according to \cite{Davison:2015bea}, the connection of $\tau$ to a momentum dissipation characteristic time 
is made up to order $\beta^2$ corrections; therefore $\tau$ can be, strictly speaking, interpreted as 
a dissipation characteristic time only when momentum is slowly dissipated \cite{Davison:2014lua} and at vanishing magnetic field. 
 This said and still sticking to $B=0$, $\tau$ proved to be a formally very useful 
quantity to furnish explicit and exact expressions for the holographic transport coefficients in all dynamical regimes \cite{Blake:2013bqa,Amoretti:2014zha,Amoretti:2014mma,Blake:2014yla}. In the present
extended context, i.e. encompassing a non-trivial magnetic field, we prefer to express everything in terms of the quantity
$h$ (as defined in \eqref{para}) both inside and outside the regime of strong momentum dissipation and independently on the magnitude of 
the magnetic field $B$. Even reminding ourselves the connection of $h$ to $\tau$ through \eqref{conne}, we adopt the former
to prevent confusion and avoid a direct hint to a momentum dissipation characteristic time
(whose explicit expression for $B\neq0$ is yet unknown).

It amounts just to a matter of algebra to verify that the transport coefficients 
derived in Section \ref{tran} can be expressed in the general form given in \eqref{aspuni}, \eqref{aspuni2}, \eqref{aspuni3}.
We highlight once more that the transport coefficients \eqref{aspuni}, \eqref{aspuni2}, \eqref{aspuni3} are expressed as functions 
of the entropy density $\mathcal{S}$, the charge density $\rho$, the magnetic field $B$, 
the conductivity $\tilde\sigma$ and $h$ only.
They \emph{formally} do not depend on any detail of the specific model that has been used to derive them.
We therefore advance the proposal that they can have wider relevance.
Within the holographic context, such a claim could be corroborated by the comparison of these 
formul\ae~with the corresponding results obtained
in other holographic models \cite{Blake:2015ina}. More generally, a careful comparison with the phenomenology
and real experiments must be pursued. In a later section we start addressing this wide question.

Finally we note that, even defining the incoherent conductivity $\sigma_Q \equiv \left[\mathcal{S}T/(\mathcal{E}+P)\right]^2$ as in \cite{Hartnoll:2007ip},
we find a mismatch between the holographic formul\ae~\eqref{aspuni}, \eqref{aspuni2}, \eqref{aspuni3} and the hydrodynamic results of \cite{Hartnoll:2007ih}. 
This is however expected, since (as recently pointed out in \cite{Davison:2015bea,Blake:2015epa}), in the zero magnetic field case 
the holographic DC formul\ae~for the thermo-electric transport agree with the modified hydrodynamic result of \cite{Hartnoll:2007ip} 
at order $1/\beta^2$ and there is a mismatch at order $\beta^0$. The discrepancy we have found here reflects most probably that an analogous 
situation arises also in the $B \ne 0$ case. To the purpose of identifying precisely the coherent and incoherent contributions
to the thermo-electric conductivities for $B \ne 0$ a careful extension of the $B=0$ analysis of \cite{Davison:2015bea} is in order.
Complementary studies could also be pursued within the fluid/gravity approach as in \cite{Blake:2015epa}.
We postpone these discussions to the future.

\subsection{Bulk electromagnetic duality and its consequences from the boundary perspective}
\label{em_duality}

The $3+1$ dimensional bulk Lagrangian \eqref{massivelag} enjoys electromagnetic self-duality. From the boundary viewpoint this
implies that the equilibrium states corresponding to two bulk solutions connected by the electromagnetic duality can be mapped
into each other, which practically means that the information regarding the thermodynamics and the transport can be interpreted
in two dual ways. Though, this does not correspond to a boundary electromagnetic duality; actually, from the holographic dictionary
it emerges clearly that the bulk electromagnetic duality exchanges the boundary magnetic field with the charge density.

The physical relevance of these duality arguments is connected to the possible description 
of the theory in terms of dual degrees of freedom and is related to the ubiquitous particle/vortex dualities of critical
or near-to-critical systems. Indeed, as noted in \cite{Hartnoll:2007ih}, in the limit $\rho, \, \tau^{-1},\, B \ll T^2$, and $\rho \sim B$ ,
the hydrodynamic transport coefficients enjoy the above mentioned duality (a priori of any gauge/gravity argument), namely, 
by exchanging the charge density with the magnetic field, $\rho \leftrightarrow B$ and the quantum critical conductivity with 
its inverse $\sigma_Q \leftrightarrow 1/\sigma_Q$, the hydrodynamical transport coefficient map into each other as follows  
\begin{eqnarray}
\label{selfdual}
&&\sigma_{xx},~\sigma_{xy},~ \alpha_{xx},~\alpha_{xy},~\overline{\kappa}_{xx},
~\overline{\kappa}_{xy} \nonumber \\
&&~~~~~~~~~~~~~~~~~\updownarrow \nonumber \\
&&\rho_{xx},-\rho_{xy}, - \vartheta_{xy},- \vartheta_{xx},~\kappa_{xx},-\kappa_{xy} \ ,
\label{dualtrans}
\end{eqnarray}
where $\hat{\rho}=\hat{\sigma}^{-1}$ is the resistivity matrix, $\hat{\theta} \equiv - \hat{\rho} \cdot \hat{\alpha}$ 
is the Nernst coefficient matrix and $\hat{\kappa}=\hat{\bar{\kappa}}-T\hat{\alpha}\cdot \hat{\rho}\cdot \hat{\alpha}$ 
is the thermal conductivity matrix at zero electric current%
\footnote{Where 
\begin{equation}
 \hat{\rho} = \rho_{xx}\, {\bm 1} + \rho_{xy}\, \hat{\epsilon}\ ,
\end{equation}
and similarly for the other transport matrices.}.

As just argued, in a gauge/gravity context, the map \eqref{selfdual} becomes 
particularly transparent as a direct consequence of the bulk electro-magnetic duality. 
So the transport coefficients \eqref{aspuni}, \eqref{aspuni2}, \eqref{aspuni3} that we have obtained holographically
naturally satisfy \eqref{selfdual} in any dynamical regime; both within and outside the hydrodynamic approximation. 
The self-duality is naturally expressed in terms of the characteristic conductivity $\tilde{\sigma}=1/q^2$
which is mapped into its inverse and not through the quantum critical conductivity $\sigma_Q$;
we remind the reader that this latter is equal to $\left[\mathcal{S}T/(\mathcal{E}+P)\right]^2$
as described in  \cite{Hartnoll:2007ip}. The electro-magnetic duality formulated in terms of $\tilde\sigma$
is exactly valid in every dynamical regime, namely
\begin{equation*}
\rho \leftrightarrow B \ , \qquad \tilde\sigma \leftrightarrow 1/\tilde\sigma \ , \, 
\end{equation*}  
\begin{eqnarray}
\label{selfdual}
&&\sigma_{xx},~\sigma_{xy},~ \alpha_{xx},~\alpha_{xy},~\overline{\kappa}_{xx},
~\overline{\kappa}_{xy} \nonumber \\
&&~~~~~~~~~~~~~~~~~\updownarrow \nonumber \\
&&\rho_{xx},-\rho_{xy}, - \vartheta_{xy},- \vartheta_{xx},~\kappa_{xx},-\kappa_{xy} \ .
\label{dualtrans}
\end{eqnarray}

The validity of \eqref{selfdual} also far from the hydrodynamic approximation has interesting consequences.
One can move away from hydrodynamical regime either lowering the temperature or increasing the magnetic field.
This statements can be made precise referring to the charge density $\rho$; lowering the temperature 
means actually to consider a regime where the ratio $T/\sqrt{\rho}$ turns to be small. Then it is possible to
appreciate that the two ways out of hydrodynamical regime just mentioned are dual in the sense of the map \eqref{selfdual}
and correspond roughly speaking to spoil criticality by means of a strong magnetic field or a strong
charge density.

\section{Holography inspired phenomenology}
\label{pheno}

In this section we discuss the possible general predictions 
based on the thermoelectric transport coefficients \eqref{aspuni}, \eqref{aspuni2}, \eqref{aspuni3} 
explicitly computed with gauge/gravity techniques.

We want to express the behavior of the holographic model \eqref{massivelag}
as independently as possible of its specific details. In line with this aim, the transport coefficients
\eqref{aspuni}, \eqref{aspuni2}, \eqref{aspuni3} were expressed in terms of
the thermodynamical quantities (computed in Section \ref{thermodynamics}) and in terms 
of the conductivity at zero net heat current $\tilde\sigma$ and the quantity $h$ defined in \eqref{para}.
These latter quantities can be thought of as ``phenomenological parameters'' whose value is not predicted 
within the model itself; as already observed, they are indeed directly related to parameters of the bulk model
($q$ and $\beta$ respectively). To clarify this idea of extending the results beyond the model used to obtain them, 
consider for example $\tilde\sigma$; for the model at hand it is related to the bulk coupling, namely  $\tilde\sigma = 1/q$. 
However, expressing all the physical results directly in terms of $\tilde\sigma$, corresponds to a model-independent 
formulation where $\tilde\sigma$ is regarded as a parameter \emph{per se} accounting for an a priori unspecified 
characteristic conductivity.

We must also recall that, as usual in bottom-up holographic models (i.e. not derived as 
consistent low-energy effective theories of UV complete string setups), we have no direct control of the microscopic 
degrees of freedom. Hence, it is particularly natural to exploit the bottom-up holographic model as a 
simple example grasping essential features of a whole class of strongly coupled theories. 
On the technical level, we rely on promoting parameters to be temperature dependent functions 
to the purpose of performing a phenomenological analysis.
As already stressed, such a logical leap constitutes a departure from the original holographic model and 
our current aim is to define how to test carefully this phenomenological approach against experimental data. When successful, it can 
inform us about universal characteristics and shed light on the mysterious behavior
of the transport properties in strongly correlated materials such as the cuprates 
(see \cite{hussey} for a wide and precise experimental report).
Previous phenomenological analyses along these lines have been performed in \cite{Blake:2014yla,Hartnoll:2014lpa,Hartnoll:2015sea,Karch:2014mba}.

The purpose of this section is to generalize to the whole set of thermo-electric transport coefficients 
the proposal of \cite{Blake:2014yla} which instead considers phenomenological scalings only for the resistivity and the Hall angle. 
There it was noted that, at low-$B$%
\footnote{The values of external magnetic fields implemented in a typical experimental set up can generally be considered small 
with respect to the intrinsic scales of the materials.}, the electric conductivity follows an inverse Matthiessen's rule: 
the conductivity at zero net heat current $\tilde\sigma$ and $\sigma_D$ (this latter directly related to 
$h$ and hence, in specific regimes, to a momentum relaxation scattering time) add up as follows
\begin{equation}
\sigma_{xx}= \tilde\sigma+\sigma_{D} \ ,\ \ \ \text{ with }\ \ \ \sigma_{D}=\frac{\rho^2 }{h} \ ,
\end{equation}
while the Hall angle $\tan \theta_H$ does not depend on $\tilde\sigma$
\begin{equation}
\tan \theta_H =\frac{\sigma_{xy}}{\sigma_{xx}} \sim \frac{B}{\rho}\; \sigma_{D} \ .
\end{equation}
In \cite{Blake:2014yla} it was also noted that, in order to fit the experimental scalings 
of the conductivity and Hall angle measured in the cuprates, namely $\rho_{xx} \sim \/\sigma_{xx}^{-1} \sim T$
and $\tan \theta_H \sim 1/T^2$, the two conductivities $\tilde\sigma$ and $\sigma_{D}$ must have 
the following scalings in temperature%
\footnote{We have chosen to express the scalings in temperature as a function of the dimensionless quantity 
$T / \sqrt{\rho}$, considering the system at fixed charge density.}
\begin{equation}
\tilde\sigma \sim \tilde\sigma^0\;  \frac{\sqrt{\rho}}{T} \ , \qquad \sigma_{D} \sim \sigma_{D}^0\; \left(\frac{\sqrt{\rho}}{T}\right)^2 \ ,
\end{equation}
where $\tilde\sigma^0$ and $\sigma_{D}^0$ are dimensionless parameter which do not depend on $T$. 
Inspired by phenomenological intuition, we have also supposed that the charge density $\rho$ 
is constant in temperature. In addition, in the quantum critical region the DC conductivity must be dominated 
by $\tilde{\sigma}$, namely $\tilde\sigma \gg \sigma_{D}$.

To make contact with both the theoretical and experimental literature, we will determine the scalings for the same transport coefficients discussed in 
\cite{Hartnoll:2015sea}, namely the the resistivity $\rho_{xx}$, the Hall angle $\tan \theta_H$, the Hall Lorentz ratio $L_{xy}$, 
the magneto-resistance $\frac{\Delta \rho}{\rho}=\frac{\rho_{xx}(B)-\rho_{xx}(0)}{\rho_{xx}(0)}$, 
the Seebeck coefficient $s=\frac{\alpha_{xx}}{\sigma_{xx}}$, the Nernst 
coefficient $\nu=\frac{1}{B}\left[\frac{\alpha_{xy}}{\sigma_{xx}}-s \tan \theta_H \right]$, 
the thermal conductivity $\kappa_{xx}$ and the thermal magneto-resistance $\frac{\Delta \kappa}{\kappa}=\frac{\kappa_{xx}(B)-\kappa_{xx}(0)}{\kappa_{xx}(0)}$.

As already argued before, we regard the transport coefficients \eqref{aspuni}, \eqref{aspuni2}, \eqref{aspuni3} as (possibly) universal 
functions of the magnetic field $B$, the charge density $\rho$, the entropy $\mathcal{S}$, 
the characteristic conductivity $\tilde\sigma$ and $\sigma_D$. 
Then, once the values of the charge density and the magnetic field are set, we need to fix the scalings 
of three quantities in order to determine the behavior of all the transport coefficients%
\footnote{Notice that this is exactly the same number of quantities which were needed to be fixed in the 
approach of \cite{Hartnoll:2015sea}.}. 
Note that the same approach cannot be followed in the hydrodynamical analysis of \cite{Hartnoll:2007ih} 
because the hydrodynamical expressions are not always writable in terms of 
$h$ which, in the small momentum dissipation and $B=0$ regime, assumes the explicit form $\frac{{\cal E}+P}{\tau}$.

In order to discuss the consequences of the proposal of  \cite{Blake:2014yla} extended to the whole set of thermoelectric transport coefficients,
we consider the following phenomenologically inspired inputs
\begin{equation}
\label{scalings}
\begin{split}
&\rho \sim \rho_0 = \text{const}  \ , \ \ \ \
\qquad \tilde\sigma \sim \tilde\sigma^0\; \frac{\sqrt{\rho_0}}{T} \ , \\
&\qquad \frac{1}{h} \sim R\; \frac{\tilde\sigma^0}{\rho_0^2}\; \left(\frac{\sqrt{\rho_0}}{T}\right)^2 \ , \ \ \ \
\qquad \mathcal{S} \sim \mathcal{S}_0\, \left(\frac{T}{\sqrt{\rho_0}}\right)^{\delta} \ ,
\end{split}
\end{equation}
where $R \equiv \frac{\sigma_{D}^0}{\tilde\sigma^0}$ and $\mathcal{S}_0$ are parameters independent of
the temperature and $\delta$ is an exponent to be phenomenologically determined.
Considering the scalings \eqref{scalings} within \eqref{aspuni}, \eqref{aspuni2} and \eqref{aspuni3}, 
in order to analyze the consequences of the proposal of \cite{Blake:2014yla},
we expand the transport coefficients at the first leading order in the dimensionless ratio $R$ 
and at weak magnetic field, namely $B/\rho_0 \ll1$. 

In order to fix the scaling exponent of the entropy density $\delta$, we compare the 
leading exponent of the thermal hall conductivity $\kappa_{xy}$ (having imposed the scalings \eqref{scalings})
with the experimental predictions for optimally doped YBCO \cite{zhang,matusiak}. 
We prefer to consider as a phenomenological input the scaling of this transport coefficient instead 
of the Hall Lorentz ratio $L_H$; the reason being that the experiments described in \cite{zhang,matusiak} 
found unexplained discrepancies both in the order of magnitude and in the temperature scaling of $L_H$ \cite{Khveshchenko:2015xea} 
while the same two experiments agree on $\kappa_{xy}$, which has to scale as $T^{-1}$. 
To reproduce this behaviour for the leading term of $\kappa_{xy}$, we have to set $\delta=1$, 
namely the proposal of \cite{Blake:2014yla} combined with the input of experimental data 
forces the entropy to scale linearly in temperature in the quantum critical region
\begin{equation}
\mathcal{S} \sim \mathcal{S}_0\ \frac{T}{\sqrt{\rho_0}} \ .
\end{equation}
This behaviour is in agreement with the experimental measurements of the electronic specific 
heat in various series of cuprates near optimal doping \cite{loram1,loram2,loram3} in a wide 
range of temperature and with the holographic analysis of \cite{Davison:2013txa}.

With this assumptions, at the first sub-leading order in the dimensionless ratio 
$R$ and at the first order in $B/\rho_0$, we find the following scalings
\begin{eqnarray}
&\rho_{xx} \sim \frac{1}{\tilde\sigma^0}\;  \frac{T}{\sqrt{\rho_0}}-\frac{\sigma_D^0}{\tilde\sigma^{0 \; 2}} \ , \\ &\frac{\Delta \rho}{\rho}\sim \tilde\sigma^0\, \sigma_{D}^0\; \left(\frac{B}{\rho_0}\right)^2\; \left(\frac{\sqrt{\rho_0}}{T}\right)^3-2 \sigma_D^{0 \;2} \; \left(\frac{B}{\rho_0}\right)^2\; \left(\frac{\sqrt{\rho_0}}{T}\right)^4 \ , \label{magre} \\
&L_{xy}\sim \frac{\mathcal{S}_0^2\, \sigma_{D}^0}{2\, \tilde\sigma^0\, \rho_0^2 }\ \frac{T}{\sqrt{\rho_0}}- \frac{5 \mathcal{S}_0^2 \sigma_D^{0 \; 2}}{4 \rho_0^2 \tilde\sigma^{0 \; 2}} \ , \\
& \tan \theta_H \sim 2\, \sigma_{D}^0\; \frac{B}{\rho_0}\; \left(\frac{\sqrt{\rho_0}}{T}\right)^2- \frac{B \sigma_D^{0 \; 2}}{\tilde\sigma^0}\; \frac{B}{\rho_0}\; \left(\frac{\sqrt{\rho_0}}{T}\right)^3 \ , \\
&\nu \sim \frac{\tilde\sigma^0}{\rho} \; \frac{\sqrt{\rho_0}}{T}-\frac{\sigma_D^0}{\rho_0} \;\left(\frac{\sqrt{\rho_0}}{T}\right)^2 \ , \label{nernst}\\
 &s \sim \frac{\mathcal{S}_0\, \sigma_{D}^0}{\rho_0\, \tilde\sigma^0}-\frac{\mathcal{S}_0 \sigma_D^{0 \; 2}}{\rho_0 \tilde\sigma^{0 \; 2}} \; \frac{\sqrt{\rho_0}}{T} \ , \label{seebeck}\\
&\kappa_{xx} \sim \frac{\mathcal{S}_0^2\, \sigma_{D}^0}{\rho_0^{3/2}}\; \frac{T}{\sqrt{\rho_0}} -\frac{\mathcal{S}_0^2 \sigma_D^{0 \; 2}}{\tilde\sigma^0 \rho_0^{3/2}}\ , \\
 &\frac{\Delta \kappa}{\kappa}\sim - (\sigma_{D}^0)^2\;\left(\frac{B}{\rho_0}\right)^2\; \left(\frac{\sqrt{\rho_0}}{T}\right)^4+\frac{2 \sigma_D^{0 \; 3}}{\tilde\sigma^0}\; \left(\frac{B}{\rho_0}\right)^2\; \left(\frac{\sqrt{\rho_0}}{T}\right)^5  \ .
\end{eqnarray}

Some of the temperature scalings derived with this approach are in accordance 
with the analysis of \cite{Hartnoll:2015sea}. There are however three discrepancies: 
the magneto-resistance for which we find a $B^2T^{-3}$ scaling instead of $B^2T^{-4}$ 
and the Nernst coefficients and the Seebeck coefficients for which the authors of \cite{Hartnoll:2015sea} 
found behaviors of the type $T^{-3/2}$ and $-T^{1/2}$ respectively. This discrepancies are due to the fact that, 
as opposed to the assumptions made in \cite{Hartnoll:2015sea}, in the present analysis the charge density is non-zero 
and the entropy has to scale linearly in temperature (instead of $T^2$ as predicted in \cite{Hartnoll:2015sea}).

Let us now make contact with the experimental literature on the cuprates. 
The Seebeck coefficient, in the normal phase of several families of cuprates,
it is usually fitted with a law of the kind $A-B\, T$ \cite{orbetelli} regardless of the doping concentration. 
In \eqref{seebeck} we found (correctly) a constant leading term while the subleading term scales as $T^{-1}$,
in contrast with the measurements of \cite{orbetelli}. Actually, in \cite{kim} deviations from the linear behaviour 
were observed at high temperature ($T\ge 300$ K); due to these deviations, a power law of the kind $T^{-1/2}$ was proposed
in \cite{Hartnoll:2015sea} but also a term $T^{-1}$ is compatible with the data. It is however not clear whether at this 
temperature the phonon-drag mechanism (which, at least in normal metals, has to be taken into account in the analysis of the thermopower)
can be neglected \cite{cohn, zhao}.

Concerning the magneto-resistance, in \cite{harris} the temperature dependence for YBCO and LSCO near optimal doping was studied. 
The result was that in YBCO $\frac{\Delta \rho}{\rho}$ follows a power law of the kind $B^2/T^n$ with $n \simeq 3.5-3.9$ while in 
LSCO a behaviour of the kind $A/(B+CT)^2$ was found. It would be interesting to make a quantitative evaluation of the unknown 
coefficient in \eqref{magre} to compute with our approach the correction to the $T^{-3}$ behaviour due to the subleading $T^{-4}$ term.

With regard to the other transport coefficients, namely $\nu$, $\kappa_{xx}$ and $\frac{\Delta \kappa}{\kappa}$,
the experimental measurements in the normal phase are not conclusive. Specifically the Nernst coefficient in \cite{wang,matusiak} 
can be seen to go to zero at high temperature in accordance to \eqref{nernst} but the data are not sufficient to determine 
a proper scaling law. Finally the thermal conductivity, and consequently also the thermal magneto-resistance, are extremely 
difficult to measure since typically in this materials the dominant contribution to these transport coefficients is given by phonons 
and not by electrons.

Except the most stable results, namely the resistivity and the Hall angle, the experiments do not seem in general conclusive 
on the other transport coefficients and we prefer to postpone any stringent comment on the possible 
connection between the present analysis and the experimental data to future discussions.

We want nonetheless to stress that, as it is evident from the formul\ae~\eqref{aspuni}, \eqref{aspuni2} 
and \eqref{aspuni3}, the behavior of the transport coefficients and that of the thermodynamical quantities are intimately related.
If the universality (at least within the class of spatially homogeneous systems) of the transport formul\ae~is confirmed, 
any proposal on the mechanism which determines the transport properties of the cuprates (at finite charge density) 
has to keep into account also the correct behavior for the thermodynamical quantities.

\label{pheno}

\section{Discussion}
\label{discu}

Relying on a membrane paradigm for holographic models featuring massive gravity, we computed analytically
all the thermoelectric transport coefficients for a strongly coupled $2+1$ dimensional system 
in the presence of a mechanism for momentum dissipation and a magnetic field perpendicular 
to the plane where the system lives. The transport coefficients can be expressed in a model-independent fashion 
in terms of the thermodynamical variables and two generic parameters accounting for 
two independent contributions to the electric conductivity. These corresponding to 
momentum conserving and momentum dissipating processes which combine to give the total electric 
conductivity according to an inverse Matthiessen's rule.

The model-independence of the expressions for the transport coefficients suggests by itself a possible general relevance 
of the formul\ae~both within and beyond the holographic realm. Inspired by this observation
(and extending the previous analysis of \cite{Blake:2014yla} to the whole set of thermoelectric coefficients), 
we enriched the holographic outcomes with phenomenological information. Specifically, we considered 
phenomenological behaviors obtained from experimental data for $\tilde\sigma$ and $\sigma_D$ (i.e. the two 
terms appearing in the inverse Matthiessen rule for the conductivity).
The set of transport coefficients supplemented with this phenomenological 
information allows us to derive the scaling properties in $T$ for all the transport quantities and compare them back 
with measurements. Furthermore, by means of comparison to measurements of the Hall Lorentz ratio, 
we fixed the scaling properties of the entropy of the system to be linear in $T$. What emerges from this 
analysis is a coherent picture which is openly exposed to be contrasted against experiments.
The definitive experimental test is however articulate and we postpone to subsequent work any 
conclusive claim.

A natural improvement of this phenomenological analysis based on the holographic formul\ae~\eqref{aspuni}, 
\eqref{aspuni2} and \eqref{aspuni3} is to keep into account the competition between different scales
and systematically analyze all the various regimes of the system. 
We refer in particular to the recent proposal of \cite{Hayes}, where, based on the experimental hint that 
in the normal phase of the cuprates \cite{Rourke,Alloul,Bourbonnais} the magneto-resistance qualitatively 
appears more $B$-linear at low temperature and more $B$-squared at high temperature, the authors suggest 
that the magnetic field and the temperature influence the transport properties in the strange metals by 
competing to set the scale of the momentum dissipation rate. Consequently they propose the following form for the momentum dissipation rate
\begin{equation}
\label{tauproposal}
\frac{1}{\tau}=\sqrt{\alpha\, T^2+ \eta\, \frac{B^2}{\rho^2}} \ ,
\end{equation}
where $\alpha$ and $\eta$ are dimensionless parameters relating the momentum dissipation rate directly to the
temperature and magnetic field respectively. In this case one has to discuss separately the weak 
$T/\sqrt{\rho} \gg B/\rho$ and strong $T/\sqrt{\rho} \ll B/\rho$ magnetic field regime. 
This constitutes one of the main future prospects of the present analysis.

As a final comment, we remind the reader that the phenomenological analysis departed from the original 
holographic model adopted for the computation of the transport coefficients. It therefore remains an open and important question 
to study the possible embedding of the dynamical promotion of the parameters into richer holographic models.

\section{Acknowledgments}
A particular thanks goes to Alessandro Braggio, Nicola Maggiore and Nicodemo Magnoli for collaboration
at an early stage of the project. A.A. wants to thank Richard Davison and Jan Zaanen for very helpful discussions.
D.M. is very grateful to Alejandra Castro, Miller Mendoza, Christiana Pantelidou and Sauro Succi for very nice and insightful discussions.
We thank the Galileo Galilei Institute for Theoretical Physics for the hospitality and the INFN
for partial support during the completion of this work. A.A. acknowledges support of a grant from the
John Templeton foundation. The opinions expressed in this publication are those of the
authors and do not necessarily reflect the views of the John Templeton foundation.

\appendix

\section{Fluctuation equations}

\subsection{Electric ansatz}
\label{El_app}

\begin{equation}
 \begin{split}
  \tilde{h}_{ti}''(z)
 &+\frac{2}{z} \tilde{h}_{ti}'(z)
  + 2 \left[\frac{\beta L^2 - B^2 z^2 \gamma^2}{L^2 f(z)} - \frac{1}{z^2}\right]\, \tilde{h}_{ti}(z)\\
 &-\frac{2 B z^2 \gamma^2 \mu }{L^2 z_h}\, \epsilon_{ij}\,\tilde{h}_{zj}(z)
  -\frac{2 \gamma^2 \mu  }{z_h}\, \tilde{a}_i'(z)
  -\frac{2 B \gamma^2}{f(z)}\, \epsilon_{ij} E_j= 0
 \end{split}
\end{equation}

\begin{equation}
 B\, \epsilon_{ij}\tilde{a}_j'(z)
 + \left(\frac{\beta}{\gamma^2} - \frac{B^2 z^2}{L^2} \right)\, \tilde{h}_{zi}(z)
 -\frac{B z^2 \mu}{z_h f(z) L^2}\, \epsilon_{ij} \tilde{h}_{tj}(z)
 +\frac{\mu}{z_h f(z)}\, E_i = 0
\end{equation}

\begin{equation}
\begin{split}
\tilde{a}_i''(z)
&+\frac{f'(z)}{f(z)}\, \tilde{a}_i'(z)
+B z\, \frac{z f'(z)+2 f(z)}{L^2 f(z)}\, \epsilon_{ij}\tilde{h}_{zj}(z)
+\frac{B z^2}{L^2}\, \epsilon_{ij}\tilde{h}_{zj}'(z)\\
&-\frac{z^2 \mu}{L^2 z_h f(z)}\, \tilde{h}_{ti}'(z)
-\frac{2 z \mu}{L^2 z_h f(z)}\, \tilde{h}_{ti}(z) = 0
\end{split}
\end{equation}

\subsection{Thermal ansatz}
\label{theapp}
\begin{equation}\label{hti_mac}
 \begin{split}
  \tilde{h}_{ti}^{(\text{th})}(z_h) = 
  \tilde{h}_{ti}^{(\text{el})}(z_h)
  &+\frac{B L^2 \epsilon_{ij} s_j \gamma ^2 \mu  \left[z_h^2 \gamma ^2 \left(B^2 z_h^2+\mu^2\right)-2 L^2 \left(z_h^2 \beta +3\right)\right]}
   {4 z_h^2 \left[-2 B^2 L^2 z_h^2 \beta  \gamma ^2+B^2 z_h^2 \gamma ^4 \left(B^2 z_h^2+\mu^2\right)+L^4 \beta ^2\right]}\\
  &\ \ \ \ -\frac{L^2s_i \left[-L^2 z_h^2 \gamma^2 \left(3 B^2 \left(z_h^2 \beta +2\right)+\beta  \mu ^2\right)\right]}
   {4 z_h^3 \left[-2 B^2 L^2 z_h^2 \beta  \gamma ^2+B^2 z_h^2 \gamma ^4 \left(B^2 z_h^2+\mu ^2\right)+L^4 \beta^2\right]}\\
  &\ \ \ \ \ \ \ \ -\frac{L^2s_i \left[B^2 z_h^4 \gamma^4 \left(B^2 z_h^2+\mu ^2\right)+2 L^4 \beta  \left(z_h^2 \beta +3\right)\right]}
   {4 z_h^3 \left[-2 B^2 L^2 z_h^2 \beta  \gamma ^2+B^2 z_h^2 \gamma ^4 \left(B^2 z_h^2+\mu ^2\right)+L^4 \beta^2\right]}
  \end{split}
\end{equation}

\begin{equation}
 \begin{split}
 \tilde{h}_{ti}''(z) 
 +\frac{2}{z}\, \tilde{h}_{ti}'(z) &
 + \left[\frac{-z^2 \gamma^2 \left(2 B^2 z_h^2+\mu ^2\right)+2 L^2 z_h^2 \beta +z^2 z_h^2 \gamma ^2 \phi'(z)^2}{L^2 z_h^2 f(z)}-\frac{2}{z^2}\right] \tilde{h}_{ti}(z) \\
&+\frac{2 B z^2 \gamma ^2 \phi'(z)}{L^2}\, \epsilon_{ij} \tilde{h}_{zj}(z) 
 +2 \gamma ^2 \phi '(z)\, \tilde{a}_i'(z) \\
&\ \ \ \ \ \ \ \ \
 +\gamma^2 \left[t\, s_i \left(\phi '(z)^2-\frac{\mu ^2}{z_h^2}\right)
 -\frac{2 B \epsilon_{ij} (E_j - s_j \phi (z))}{f(z)}\right] = 0
 \end{split}
\end{equation}

\begin{equation}
 \begin{split}
  4 B L^2 z^3 z_h^2 & \gamma^2 f(z)\, \epsilon_{ij} \tilde{a}_j'(z)
  +4 z_h^2 f(z) \left(L^2 z^3 \beta -B^2 z^5 \gamma ^2\right)\, \tilde{h}_{zi}(z)\\
 &-4 B z^5 z_h \gamma ^2 \mu \, \epsilon_{ij} \tilde{h}_{tj}(z)
  -6 L^4 z_h^2 s_i f(z)
  +2 L^4 z_h^2 s_i \left(z^2 \beta +3\right) \\
 &\ \ \ \ \ \ \ \ \ \ \ \ \ \ 
  +L^2  z^3 \gamma ^2 \left[-B^2 z z_h^2 s_i + 4 E_i z_h \mu + s_i \mu^2 (3 z - 4 z_h)\right] = 0
 \end{split}  
\end{equation}

\begin{equation}
 \begin{split}
   \tilde{a}_i''(z)
  &+\frac{z^4 \gamma ^2 \left(B^2 z_h^2+\mu^2\right)+6 L^2 z_h^2 f(z)-2 L^2 z_h^2 \left(z^2 \beta +3\right)}{2L^2 z z_h^2 f(z)}\, \tilde{a}_i'(z) \\
  &+ B z \frac{z^4 \gamma ^2 \left(B^2 z_h^2+\mu ^2\right)+10 L^2 z_h^2 f(z)-2 L^2 z_h^2 \left(z^2 \beta+3\right)}{2 L^4 z_h^2 f(z)}\, \epsilon_{ij} \tilde{h}_{zj}(z) \\
  &\ \ \ \ \ \ \
   +\frac{B \epsilon_{ij} s_j}{f(z)}
   +\frac{B z^2}{L^2}\, \epsilon_{ij} \tilde{h}_{zj}'(z)
   -\frac{z^2 \mu }{L^2 z_h f(z)}\, \tilde{h}_{ti}'(z)
   -\frac{2 z \mu}{L^2 z_h f(z)} \tilde{h}_{ti}(z) = 0
 \end{split}
\end{equation}

\subsection{Stress-energy tensor with thermal gradient}
\label{appthermalT}

In this appendix we want to clarify a very subtle aspect of the computation of the DC transport coefficient illustrated 
in \cite{Donos:2014cya}, namely the fact that the stress energy tensor $T^{ti}$ \eqref{Tti} and the quantity 
$\frac{\sqrt{-g}}{\kappa_4^2}\nabla^z k^i$ coincide once evaluated on the boundary up to a term linear in the time coordinate $t$;
this latter however does not contribute to the DC response as clearly explained in Appendix C of \cite{Donos:2014cya}.

Once evaluated on the thermal ansatz \eqref{ansaflu2}, the two quantities previously mentioned take the following form
\begin{equation}
\label{ttt}
\frac{\sqrt{-g}}{\kappa_4^2}\nabla^z k^i=\frac{1}{\kappa_4^2}\left(\tilde{h}_{ti} \left(\frac{f(z)}{z}-\frac{f'(z)}{2}\right)+\frac{f(z)}{2}\tilde{h}_{ti}'\right) \ ,
\end{equation}
\begin{multline}
\label{ttt1}
 T^{ti} = \frac{L^5}{\kappa_4^2 z^5} \left(-K^{ti} + K g^{ti}_b + \frac{2}{L}g^{ti}_b\right)=\\
        \frac{\tilde{h}_{ti}'(z)}{2 \kappa_4 ^2 \sqrt{f(z)}}-\frac{\tilde{h}_{ti}(z)}{z \kappa_4^2 \sqrt{f(z)}}+\frac{2\, \tilde{h}_{ti}(z)}{z \kappa_4 ^2 f(z)} \\
        +t \alpha_i\left(- \frac{2 L^2}{z^3 \kappa_4^2}+\frac{2 L^2 \sqrt{f(z)}}{z^3 \kappa_4^2}-\frac{L^2 f'(z)}{2 z^2 \kappa_4^2 \sqrt{f(z)}} \right)\ .
\end{multline}
In order to evaluate the previous quantities at the boundary $z=0$ we substitute the background solution \eqref{solublack} for $f(z)$ and we 
impose $\tilde{h}_{ti} \sim h_{ti}^{(0)} z$ near $z=0$. This latter condition is due to the fact that, as explained in \cite{Donos:2014cya}, 
we need to switch off the term proportional to $z^{-2}$ in the asymptotic expansion for $\tilde{h}_{ti}$ in order to avoid additional thermal 
deformation associated to this mode. Keeping into account these asymptotic behaviors it is easy to verify that the $z \rightarrow 0$ limit 
of \eqref{ttt} coincides with the time independent part of \eqref{ttt1} evaluated in the same limit.

\end{document}